\documentclass[prl,twocolumn,floats,aps,epsfig,nofootinbib,axodraw,amssymb]{revtex4}
\usepackage{subfigure}
\def\beq{\begin{equation}}
\def\eeq{\end{equation}}
\def\bea{\begin{eqnarray}}
\def\eea{\end{eqnarray}}

\def\gev{\rm GeV}
\def\tev{\rm TeV}

\def\tt{t\bar t}
\def\etmiss{E\!\!\!\!\slash_{T}}
\def\pslash{\not{\hbox{\kern-4pt $p$}}}
\def\qslash{\not{\hbox{\kern-4pt $q$}}}
\def\lv{\not{\hbox{\kern-4pt $L$}}}
\def\lsim{\mathrel{\raise.3ex\hbox{$<$\kern-.75em\lower1ex\hbox{$\sim$}}}}
\def\gsim{\mathrel{\raise.3ex\hbox{$>$\kern-.75em\lower1ex\hbox{$\sim$}}}}
\def\ifmath#1{\relax\ifmmode #1\else $#1$\fi}

\usepackage{graphicx}
\usepackage{bm}
\begin{document}
\draft
\renewcommand{\thefootnote}{\arabic{footnote}}

%



\title{Top Quark Pairs at High Invariant Mass -- \\  A Model-Independent Discriminator of New Physics at the LHC}
\bigskip
\author{Vernon Barger, Tao Han and Devin G. E. Walker\footnote{Email Address: barger, than, walker@physics.wisc.edu.}}
\address{Department of Physics, University of Wisconsin, Madison, WI 53706, U.S.A.}

\begin{abstract}
We study top quark pair production to probe new physics at the LHC. We propose reconstruction methods for  $t\bar{t}$ semileptonic events and use them to reconstruct the  $t\bar{t}$ invariant mass. The angular distribution of top quarks in their c.m. frame can determine the spin and production subprocess for each new physics resonance. Forward-backward asymmetry and CP-odd variables can be constructed to further delineate the nature of new physics. We parametrize the new resonances with a few generic parameters and show high invariant mass top pair production may provide an early indicator for new physics beyond the Standard Model. 
\end{abstract}

\maketitle

 In the next few years, high energy physics will experience the excitement of major discoveries when the CERN Large Hadron Collider (LHC) opens up the unexplored TeV energy scale. Besides the long anticipated Higgs boson that is responsible for the mass generation in the highly successful Standard Model (SM), the quadratic sensitivity of the Higgs boson mass to radiative corrections indicate the existence of new physics associated with electroweak symmetry breaking naturally at the scale of order $4\pi v$.\footnote{Here $v\approx 246$ GeV is the Higgs field vacuum expectation value.}
 Numerous extensions to the Standard Model (SM) have been proposed to describe electroweak symmetry breaking.  A sample of popular scenarios include:  the Minimal Supersymmetric Standard Model (MSSM) \cite{Dimopoulos:1981zb}, new strong dynamics \cite{technicolor,topcolor,topseesaw,tcreview}, composite Higgs at the TeV scale \cite{Kaplan:1983sm}, Little Higgs theory \cite{LH}, and extra dimensions at the electroweak 
 scale \cite{Arkani-Hamed:1998rs,Randall:1999ee}.  String-inspired extensions in the gauge sector associated with an extra  $U(1)$ symmetry \cite{Zprime} also lead to striking signatures.  It is therefore highly expected that many new signatures will become manifest at TeV energy scales that can be probed at the LHC.   

The LHC will be a ``top factory":  About 80 million $t\bar{t}$ events will be generated by QCD production with an 
integrated luminosity of 100 fb$^{-1}$.  Thus studying top-quark production can be fruitful.  The fact that the top quark mass is at the electroweak scale ($m_t\approx v/\sqrt 2$)  suggests that top-quark production may be sensitive to new physics near the TeV scale.  Generically, if the new physics contributes to $\tt$ production as an $s$-channel 
resonance, we want to identify the signal as a bump on the smoothly falling $t\bar t$ invariant mass  distribution. Once we can reconstruct the $\tt$ c.m.~frame, the integer spin ($J= 0, 1, 2$) of any resonances can  be determined from the polar
angular  distribution of the top quark.  An asymmetry of this distribution would probe the chiral structure of the couplings. It may be possible to explore the CP property of the couplings with the help of CP-odd kinematical variables constructed from the final state particle momenta. Moreover, the relative importance of gluon-gluon and quark-antiquark subprocesses can be inferred from the spin and angular distributions.  It is thus of fundamental importance  to effectively reconstruct the $\tt$ invariant mass via their decay products. 

We focus on the semileptonic decay mode, $ \tt \to bj_1j_2\ \bar b\ell^- \bar\nu + \mathrm{c.c.}$ where
$\ell=e$ or $\mu$.
The purely hadronic decay mode of $\tt$ not only suffers from a much larger QCD background, but also loses the identification of $t$ from $\bar t$.
For the purely leptonic mode, with a small  branching fraction of about $4/81$, 
one cannot reconstruct the $\tt$ invariant mass with two missing neutrinos.  
The signal to search for is an isolated charged lepton plus missing energy 
($\etmiss$), 2 $b$-jets plus 2 light jets. The branching ratio of  the semileptonic to the hadronic channel is 2/3.

\vskip 0.2cm
\noindent
\underline{New Reconstruction Methods}:

A primary focus of our study is to reliably reconstruct $t\bar{t}$ kinematics at high invariant mass on an event-by-event basis.  The challenge is to reconstruct the momentum of the missing neutrino.  The transverse momentum  of the neutrino is identified with the observed $\etmiss$.  The longitudinal momentum  is subject to a two-fold ambiguity from solving the kinematic quadratic equation. 

Several top reconstruction methods have been used at the Tevatron \cite{tevatron}.  There, however, the top quarks are produced near threshold and the kinematics of the subsequent decay products are very complicated. Since we are interested in new physics in the TeV region, demanding a high invariant mass for the $\tt$ events  will tremendously
simplify the kinematics, especially by distinguishing the $b$ quark from $\bar b$.  Throughout our study, we use a $2 \to 6$ partonic level  monte-carlo simulation that incorporates full spin correlations from production through decay \cite{matrixelements}. 
 We made a Pythia simulation, including gluon radiation and hadronization, 
 that confirmed our basic results.  

We impose a cluster transverse mass cut on the $\tt$ system
\begin{eqnarray*}
M_T= \sqrt{(p_{b} + p_{\bar b} + p_{j_1} + p_{j_2} + p_\ell)^2 + \etmiss^2} + \etmiss > 600\ {\rm GeV}.
\label{mtc}
\end{eqnarray*}
We adopt kinematical cuts from the ATLAS and CMS \cite{atcmscuts} top studies.  We smear the hadronic energy according to a Gaussian error given by 
$\Delta E_j/E_j=0.5/\sqrt {E_j/\gev} \oplus 0.03$;  
and the lepton momentum by 
$\Delta p_T^\ell /p_T^\ell= 0.36 (p_T^\ell/\tev) \oplus 0.013/\sqrt{\sin\theta}$,
where $\theta$ is the polar angle of the lepton with respect to the beam direction in the lab frame.  
We present two schemes to reconstruct semileptonic $t\bar{t}$ events and evaluate their efficacy.

\vspace{0.2cm}
\noindent
 (1). $(M_W, m_t)$ {\it  scheme}: \\
   In this scheme, the key assumption is to take $M_W$ and $m_t$ as inputs for their on-shell production and decays. \\
 {\it Step I:} 
 Demand $m_{l \nu}^2 = M_W^2$.  The longitudinal momentum of the neutrino is formally expressed as
\begin{displaymath}
p_{\nu L}  ={1 \over {2\, p_{e T}^2}}
 \left( {A\, p_{e L} \pm E_e \sqrt{A^2  - 4\,{p}^{\,2}_{e T}\etmiss^2}} \right),
\end{displaymath}
where $A = M_W^2 + 2 \,\vec{p}_{e T} \cdot \,\vec{\etmiss} $.  If $A^2 - 4 \, {p}^{\,2}_{e T}\,\etmiss^{\,2} \geq 0$, 
the value of $p_{\nu L}$ that best yields the known top mass via $m_{l\nu b}^2 = m_t^2$ is selected.  This ideal situation may not always hold when taking into account the detector resolutions.  
For cases with no real solutions, we then  proceed to the next step.\\
 {\it Step II:} To better recover the correct kinematics, 
we instead first reconstruct the top quark directly by demanding $m_{l\nu b}^2 = m_t^2$.  
The longitudinal momentum of the neutrino is expressed as
\begin{eqnarray*}
p_{\nu L} & = &  {A'\, p_{bl L}}/ 2(E_{bl}^2 - p_{bl L}^2) \  \pm  {1\over 2(E_{bl}^2 - p_{bl L}^2)}\\
&\times & ( {{p_{bl L}^2 A^{'\,2}  + (E_{bl}^2 - p_{bl L}^2) \,
( A^{'\,2} - 4 E_{bl}^2 E\!\!\!\!\slash_{T}^2  )} })^{1/2} ,
\end{eqnarray*}
where $A' = m_t^2 - M_{bl}^2 + 2 \,\vec{p}_{bl T} \cdot \,\vec{\etmiss} $.
The two-fold ambiguity is broken by choosing the value that best reconstructs 
$M_W^2 = m_{l \nu}^2 $. 
A plot of the top and $W$ mass distributions is  shown in  Fig.~1(a).  
The  solid histogram is from the procedure {\it Step I},  and  
the  dashed histogram from {\it Step II}. 
With these two steps, there could still be some events that
do not lead to a real solution, We thus discard them in our event collection. 
The discard rate is about 16\%.

\begin{figure}
{\label{fig:mwmt}
	\includegraphics[width=7truecm,height=6truecm,clip=true]{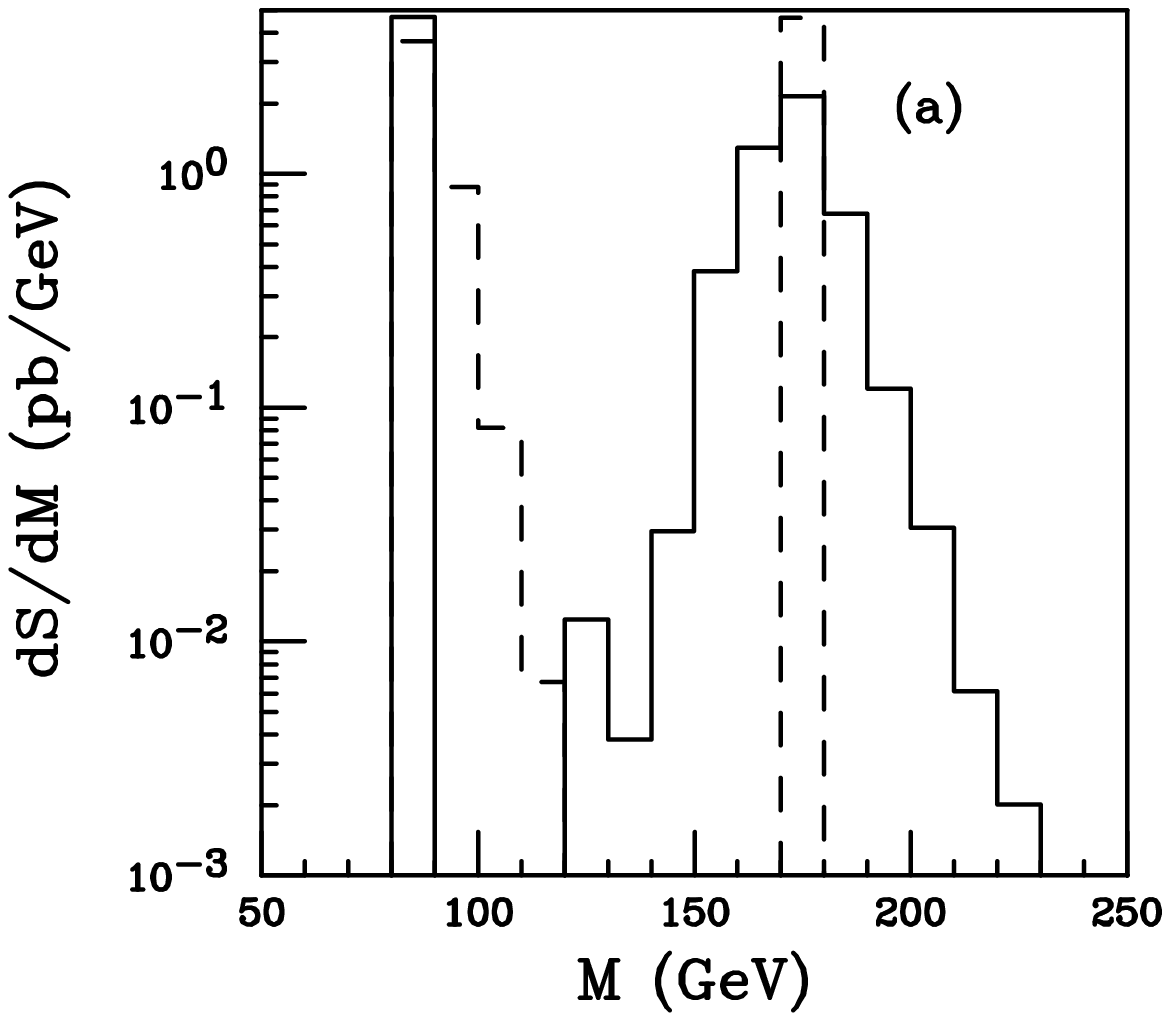}}
{\label{fig:3Dangle}	
	\includegraphics[width=7truecm,height=6truecm,clip=true]{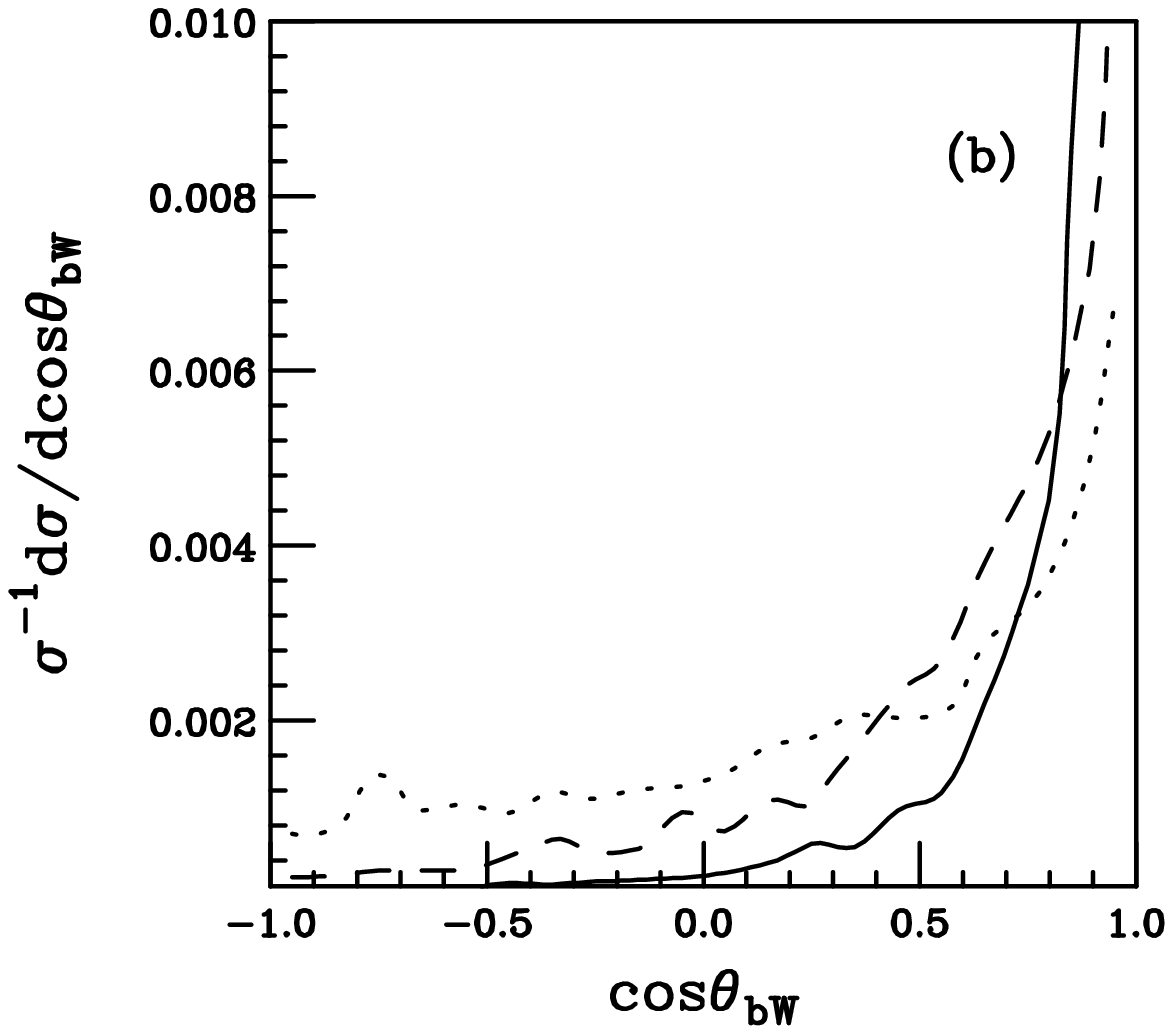}}
\caption{ (a) The $W$ and top mass reconstructions from the ($M_W$, $m_t$) scheme, 
with the procedure Step I (solid) and Step II (dashed).
(b) Differential cross section versus $\cos{\theta}_{b\,W}$ with no invariant mass cut (dotted),
and with cuts of 600 GeV (dashed) and 1000 GeV (solid). }
\end{figure}
 
\vskip 0.3cm
\noindent
{(2). \it Small $\theta_{bW}$ angle selection scheme}:\footnote{We have chosen to use $\theta_{bW}$ in our reconstruction instead of, e.g., $\theta_{l\nu}$, because the $b$ quark is on average much more energetic (highly boosted) than the lepton.}\\ 
This scheme reconstructs the $t\bar{t}$ system without relying on the top mass reconstruction, thus avoiding potentially large QCD corrections due to gluon radiations.  Since we are interested in new physics in the TeV regime, the top quarks will be relativistic with a $\gamma$-factor of (1 TeV)$/2m_t \sim 3$. We thus expect that the decay products are fairly collimated along the top quark moving direction. This is illustrated in Fig.~1(b) for the normalized opening-angle distribution between $b$ and $W^+$ in the lab frame,  
where an increasing  cluster transverse mass cut has been imposed for the dotted, dashed and solid curves for $M_{T} >0,\  600,\ 1000$ GeV, respectively.  With the initial requirement $m_{l\nu}^2 =M_W^2$, the two-fold ambiguity in $\theta_{bW}$ is resolved by choosing the smaller reconstructed angle. This scheme should work better at  higher $\tt$ invariant masses.
 
\begin{figure}
{
\includegraphics[width=7truecm,height=6truecm,clip=true]{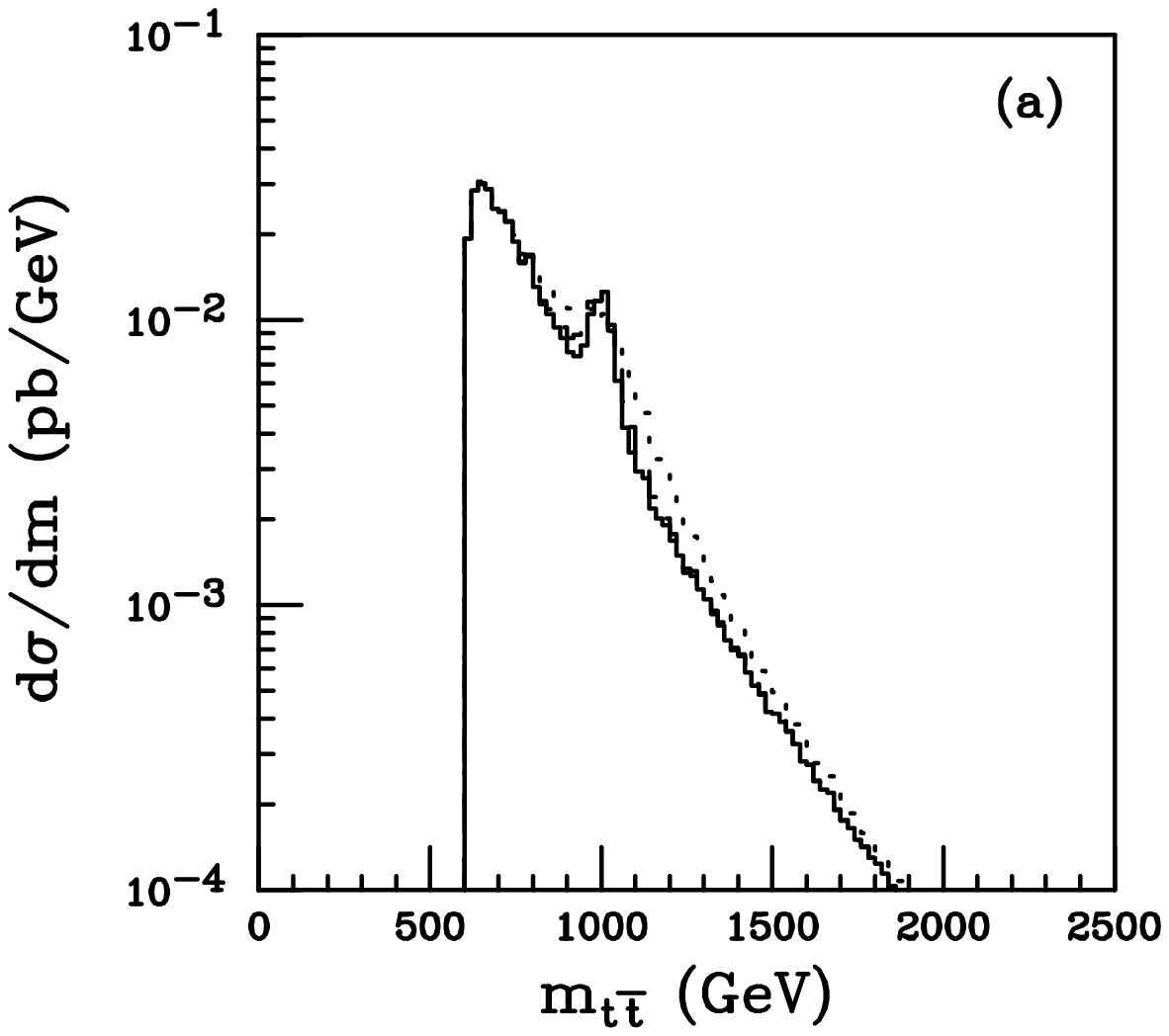}}
{	
\includegraphics[width=7truecm,height=6truecm,clip=true]{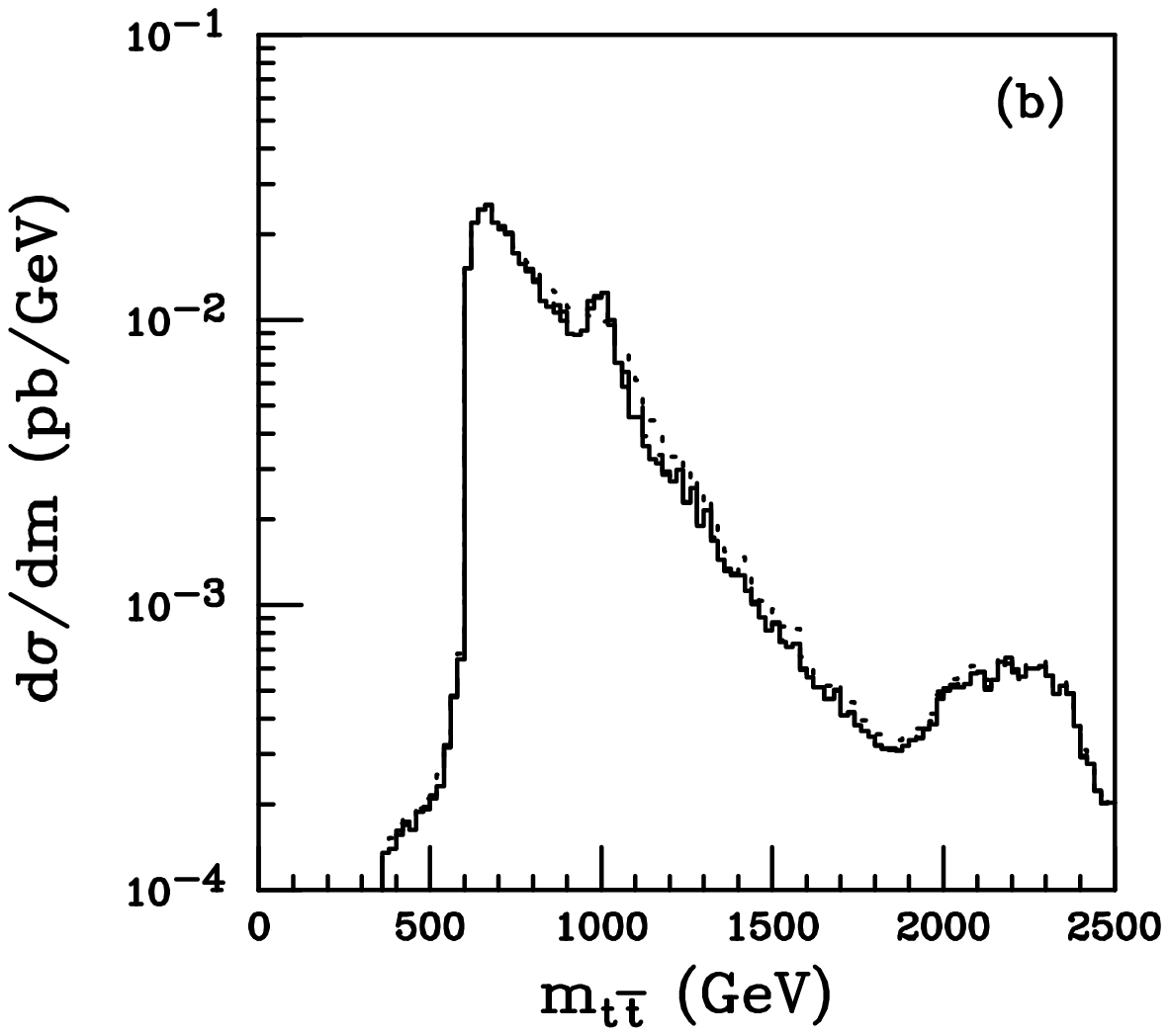}}
\caption{$t\bar{t}$ invariant mass distributions reconstructed from (a) the ($M_W$, $m_t$) scheme, and (b) the small angle selection scheme.  Both plots featured a 1 TeV resonance with a total width of 2\% (solid), 5\% (dashed), and 20\% (dotted) of the resonance's mass. }
\label{Fig:LHC}
\end{figure}

\vskip 0.2cm
\noindent
\underline{Backgrounds to the $t \bar t$ Signal}:

The major backgrounds to our $t\bar t$ events include the processes $W+$ jets, $Z+$ jets, $WW$, $WZ$ and $ZZ$.  The ATLAS and CMS Technical Design Reports \cite{tdr} detail studies of the selection efficiencies for these background processes in comparison to a reconstructed $t\bar{t}$ semileptonic signal.  The ATLAS (CMS) group found for an integrated luminosity of $10$ fb$^{-1}$ ($1$ fb$^{-1}$)  a signal to background ratio of $S/B = 65$ ($S/B = 26$) \cite{atcmscuts}.  Because of the expected high $S/B$ ratio,  our analysis is concentrated solely on the $\tt$ events without including the small background contamination.  Our analysis does not include misidentification of faked leptons from jets in $t\bar{t}$ total hadronic decays.

Although the $t$ ($\bar{t}$) is primarily identified by the charged lepton, $\ell^+$ ($\ell^-$), a concern is the matching of the b-jet associated with top quark decay.  Both ATLAS and CMS studies  \cite{tdr} show a combination of kinematic fits, designed to properly reconstruct the $W$ boson and the hadronically decaying top significantly reduces misidentification.  Our cut on $M_T$ helps significantly in this regard.  

\vskip 0.2cm
\noindent
\underline{Search for New Physics}:

We wish to explore the new physics searches in a model-independent manner 
for $t\bar{t}$ semi-leptonic decays.  We consider $t\bar{t}$ production via
\begin{eqnarray*}
g  g \to \,\,\phi \to t \bar{t},    \quad
q  \bar q \to \,\,V \to t \bar{t},   \quad
q \bar q,\  gg  \to \,\,\tilde{h} \to   t \bar{t},
\end{eqnarray*}
where $\phi$, $V$ and $\tilde{h}$ are the spin-0, spin-1, and spin-2 resonances.  We characterize the effects on the invariant mass spectrum with three parameters:  mass, total width, and 
the signal cross section normalization  ($\omega^2$).  The normalization $\omega=1$ defines our benchmark for the spin 0, 1 and 2 resonances.  They correspond to the SM-like Higgs boson, a $Z'$ with electroweak coupling strength and left (L) or right (R) chiral couplings to SM fermions, and the Randall-Sundrum graviton $\tilde h$ with the couplings scaled as
$\Lambda^{-1}$ for $\tilde h q\bar q$, and $(\Lambda \ln(M^*_{pl}/\Lambda))^{-1}$ for $\tilde h gg$, respectively.\footnote{More precisely, we use the Feynman rules given in \cite{HLZ} and include the additional
warp correction factors from \cite{DHR}.}  Numerically, we take $\Lambda = 2$ TeV.

In Fig.~2 we show the reconstructed $t\bar t$ invariant mass distributions for the two reconstruction schemes.  The SM $t\bar{t}$ total cross section is theoretically known beyond the leading order in QCD \cite{QCDcorrect}.  We thus expect to have a good control of this distribution even at high invariant masses.  As for new physics,  we include the contribution of  a 1 TeV vector resonance for illustration, for $\omega_\mathrm{v} = 1$, with total widths specified in the caption of Fig~2.  We note that a very high invariant mass tail exists for the $t\bar{t}$ invariant mass reconstructed via the small angle selection.  This comes from  the mis-reconstructed events in the low invariant mass region.  When a large enough transverse mass cut is applied for a given resonance, the tail will not obfuscate the resonance signal.  For example, if searching for 2 TeV resonance, a 800 GeV minimum cut will eliminate the tail for the mass region of interest.

We maximize the signal observability by isolating the resonance within an invariant mass window of $\pm 100$~GeV, $\pm 30$~GeV and $\pm 25$~GeV for the scalar, vector and graviton resonance, respectively.  Given a resonance mass and total width, we can  quantify how large $\omega$ needs to be for a $5 \sigma$ discovery.  With the number of events for a signal (S) and background (B), we require $S/\sqrt{B+S} > 5$.  This translates to a bound $\omega^2 > ( 25 + 5 \sqrt{25 + 4 B}) / 2S_1$ where $S_1$ is the benchmark signal rate for $\omega=1$.  This is illustrated by Fig.~3 versus the mass for a scalar, vector and graviton resonance for total widths of 20\%, 5\%, and 2\% of its mass,  respectively, for an integrated luminosity of 10 fb$^{-1}$.

\begin{figure}
{\includegraphics[width=7truecm,height=6truecm,clip=true]{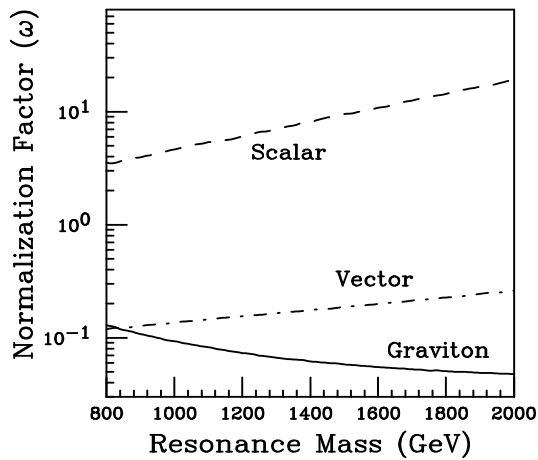}}
\caption{Normalization factor versus the resonance mass 
for the scalar (dashed) with a width-mass ratio of $20\%$, 
vector (dot-dashed) with 5\%,  and graviton (solid) 2\%, respectively.  
The region above each curve represents values of $\omega$ that give 5$\sigma$ 
or greater statistical  significance with 10 fb$^{-1}$ integrated luminosity.}
\label{fig:parameterscan}
\end{figure}

It is of critical importance to reconstruct the c.m.~frame
of the resonant particle, where the fundamental properties
of the particle can be best studied.  In Fig.~4, we show the top quark angular distribution, 
$\cos\theta^*$, with $\theta^*$ defined as the angle  in the $t\bar t$ c.m.~frame 
between the top-quark momentum and the incident quark momentum, 
with latter determined by the longitudinal boost direction of the c.m.~system.  
Although events in the forward and backward regions are suppressed due to the stringent kinematical cuts, we still see the impressive features of the $d$-function distributions \cite{PDG} 
in Fig.~4(a):  a flat distribution for a scalar resonance (dashed), 
$d^1_{11}$ distribution for the left/right chiral couplings of a vector (dotted), 
and $d^2_{1\pm1}$ from $q\bar q$ (solid) and 
$d^2_{2\pm 1}$ from $gg$ (dot-dashed) for a spin-2 resonance.
To illustrate the statistical sensitivity for observing the characteristic distributions, we
show in Fig.~4(b) the expected SM $t\bar t$ events (solid)  with 1$\sigma$ statistical
error bars in each bin for a 10 fb$^{-1}$ integrated luminosity, along with a $5\sigma$ signal
of a chirally coupled vector summed with the $t\bar t$  background in the resonant region (dashed). 
Due to the large event sample, the statistical significance is evident in the central and forward
region. 
The forward-backward asymmetry in $\cos\theta^*$ 
can thus be constructed to probe the chiral couplings of the particle to the top quark. 
With the identification of the charged leptons, one may even form kinematical 
triple products to test the CP properties of the couplings \cite{Darwin}. As for distinguishing 
the other spin states, it is a question of statistical significance. A narrow graviton may be
relatively easy to confirm, achievable with a few tens of fb$^{-1}$ luminosity; 
a broad scalar may require the highest luminosity, more than $300$ fb$^{-1}$.

\begin{figure}
{\includegraphics[width=7.5truecm,height=6truecm,clip=true]{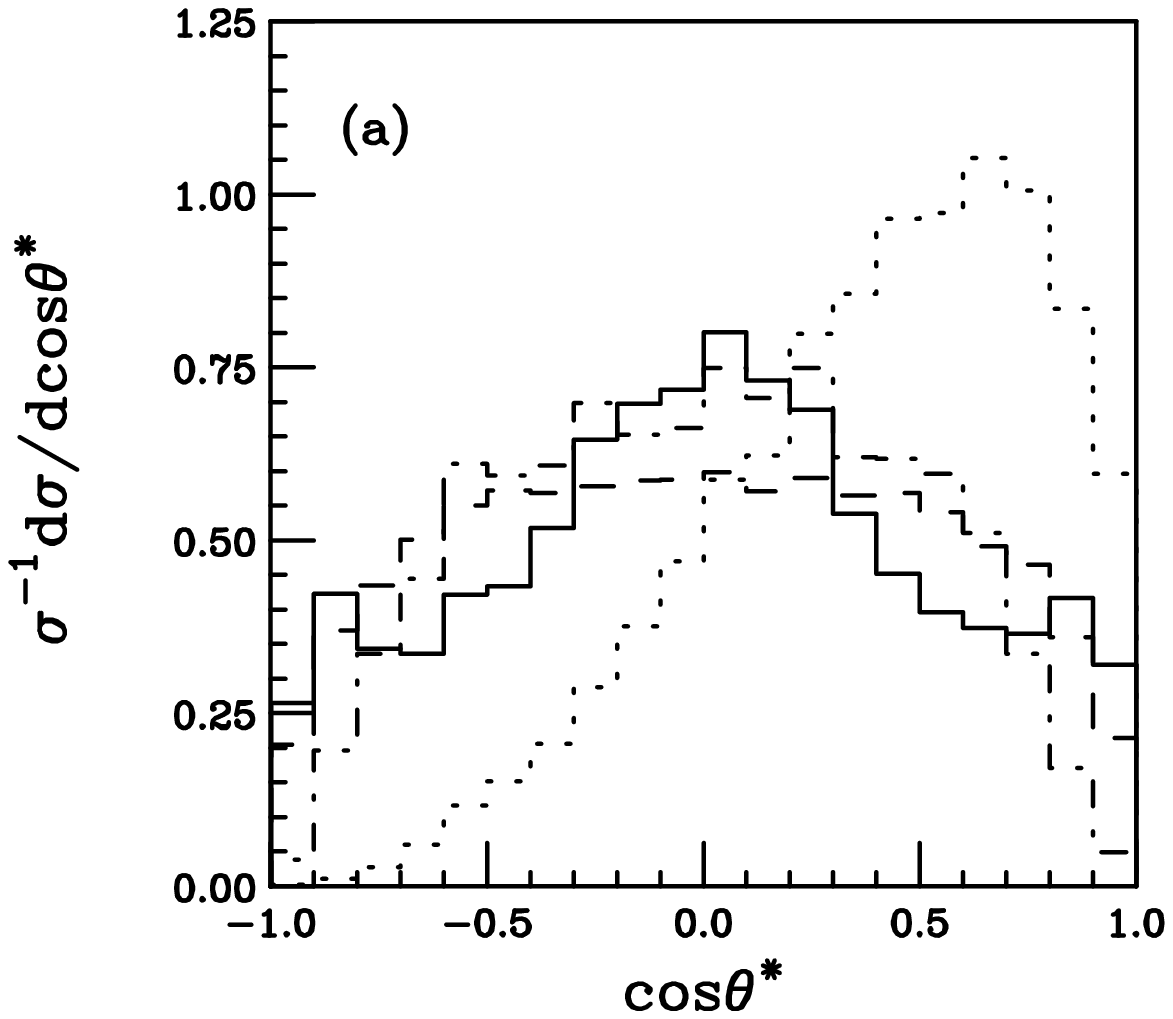}}
{\includegraphics[width=7.5truecm,height=6truecm,clip=true]{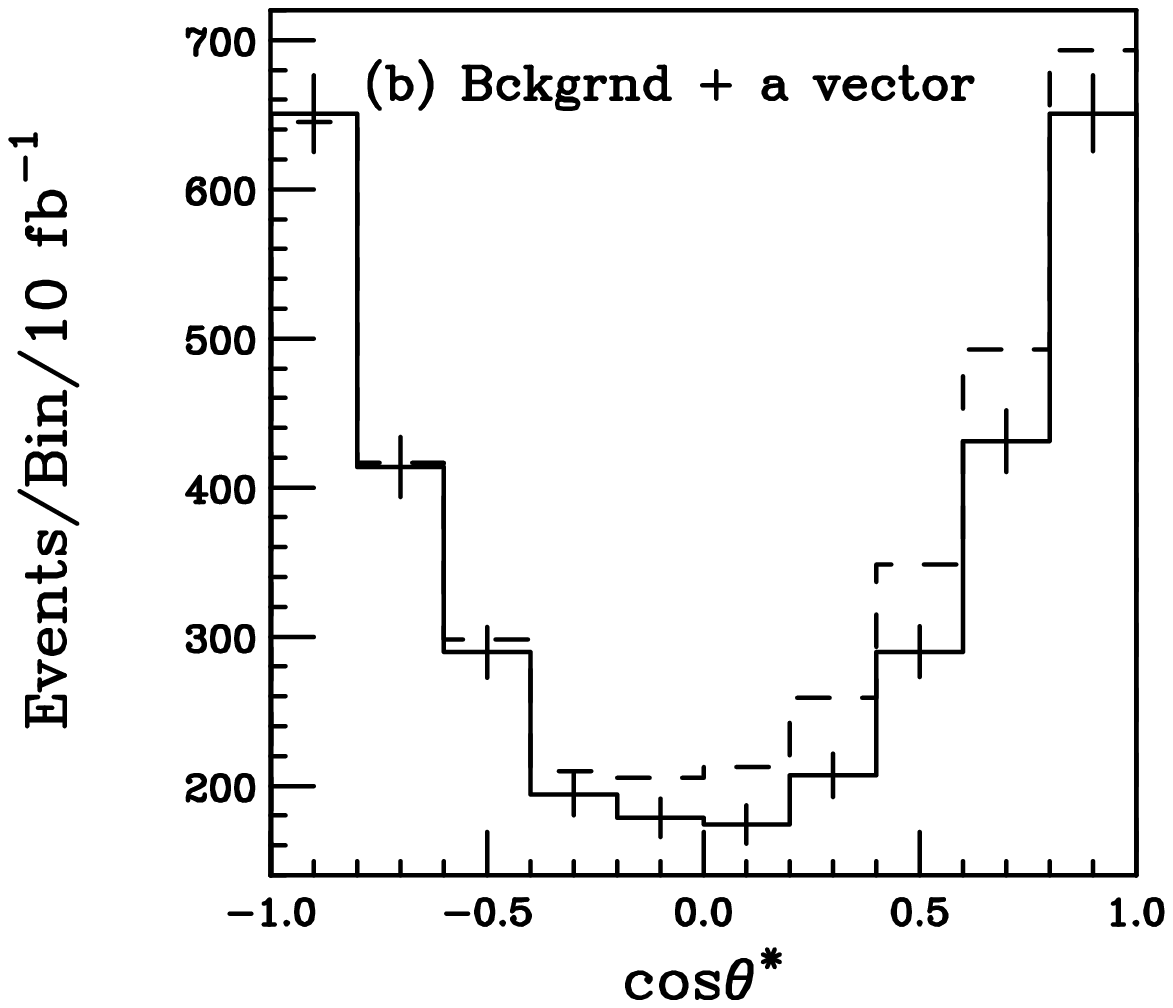}}
\caption{Polar angular distributions for the top quark in the c.m.~frame, 
(a) Signal only by the ($M_W$, $m_t$) scheme 
for a scalar (dashed), a vector (dots), and a graviton from $q\bar q$ (solid)
or from $gg$ (dot-dashed);
(b) number of events for the SM $t\bar t$ background (solid) with 
1$\sigma$  statistical error bars, 
and the background plus a vector resonance (dashed).}
\label{fig:angledist}
\end{figure}

In summary, we investigated two ways to reconstruct semileptonic $t\bar{t}$ events at high $t\bar{t}$ invariant mass and showed the utility of each in discovering new physics in the form of integer-spin resonances.  The angular distributions of the top in the reconstructed CM frame reveal the spin of the resonance, and relative contribution from the initial states $q\bar q$ or $gg$.  The forward-backward asymmetry and CP-odd variables can be constructed to further differentiate models.  Because SM top quark physics is well predicted, high invariant mass top pair production may provide an early indicator for new physics beyond the Standard Model at the LHC.

\vskip 0.2cm
\noindent
{\it Acknowledgments:}
We thank M.~Franklin, M. Herndon, J.~Hewett, I.~Hinchcliffe, G.-Y.~Huang, F.~Petriello, T.~Rizzo, W.~Smith, M.~Spalinski, and L.-T. Wang for helpful discussions.  This work was supported in part by the U.S.~Department of Energy
under grant DE-FG02-95ER40896 and the Wisconsin Alumni Research Foundation.


\begin{thebibliography}{000}
\bibitem{atcmscuts}
Following ATLAS and CMS, we adopt the kinematical cuts as 
$E_T^j > 20$ GeV, $|\eta_j | < 2.5$;  
$p_T^\ell > 20$ GeV, $|\eta_\ell| < 2.5$, and $\etmiss >20$ GeV.
The lepton-jet and jet-jet isolation is $\Delta R > 0.4$.
\end{thebibliography}
\end{document}